\documentclass{article}

\usepackage{arxiv}

\usepackage{microtype}
\usepackage{graphicx}
\usepackage{booktabs} 
\usepackage{amsmath} 
\usepackage{mathtools}
\usepackage{caption}
\usepackage{subcaption}
\usepackage{natbib}
\usepackage{amssymb}
\usepackage{algorithm}
\usepackage{algorithmic}

\usepackage[colorinlistoftodos,textsize=tiny,textwidth=18mm]{todonotes}

\usepackage{hyperref}

\makeatletter
\newcommand{\printfnsymbol}[1]{%
  \textsuperscript{\@fnsymbol{#1}}%
}
\makeatother

\title{Training Data Leakage Analysis in Language Models}

\author{{Huseyin A. Inan\thanks{Equal Contribution}}\\
	Microsoft Research\\
	\texttt{huseyin.inan@microsoft.com} \\
	\And
	{Osman Ramadan\printfnsymbol{1}} \\
	Microsoft Corporation\\
	\texttt{osman.ramadan@microsoft.com} \\
	\And
	{Lukas Wutschitz} \\
	Microsoft Corporation\\
	\texttt{lukas.wutschitz@microsoft.com} \\
	\And
	{Daniel Jones} \\
	Microsoft Corporation\\
	\texttt{t-dajon@microsoft.com} \\
	\And
	{Victor Rühle} \\
	Microsoft Corporation\\
	\texttt{virueh@microsoft.com} \\
	\And
	{James Withers} \\
	Microsoft Corporation\\
	\texttt{jawithe@microsoft.com} \\
	\And
	{Robert Sim} \\
	Microsoft Research\\
	\texttt{rsim@microsoft.com} \\
}

\date{}

\renewcommand{\headeright}{}
\renewcommand{\undertitle}{}

\begin{document}
\maketitle

\begin{abstract}
Recent advances in neural network based language models lead to successful deployments of such models, improving user experience in various applications.
It has been demonstrated that strong performance of language models comes along with the ability to memorize rare training samples, which poses serious privacy threats in case the model is trained on confidential user content.
In this work, we introduce a methodology that investigates identifying the user content in the training data that could be leaked under a strong and realistic threat model.
We propose two metrics to quantify user-level data leakage by measuring a model's ability to produce unique sentence fragments within training data.
Our metrics further enable comparing different models trained on the same data in terms of privacy.
We demonstrate our approach through extensive numerical studies on both RNN and Transformer based models.
We further illustrate how the proposed metrics can be utilized to investigate the efficacy of mitigations like differentially private training or API hardening.
\end{abstract}

\section{Introduction}
\label{sec:intro}
\begin{figure}[ht]
\vskip 0.2in
\begin{center}
\centerline{\includegraphics[width=.7\columnwidth]{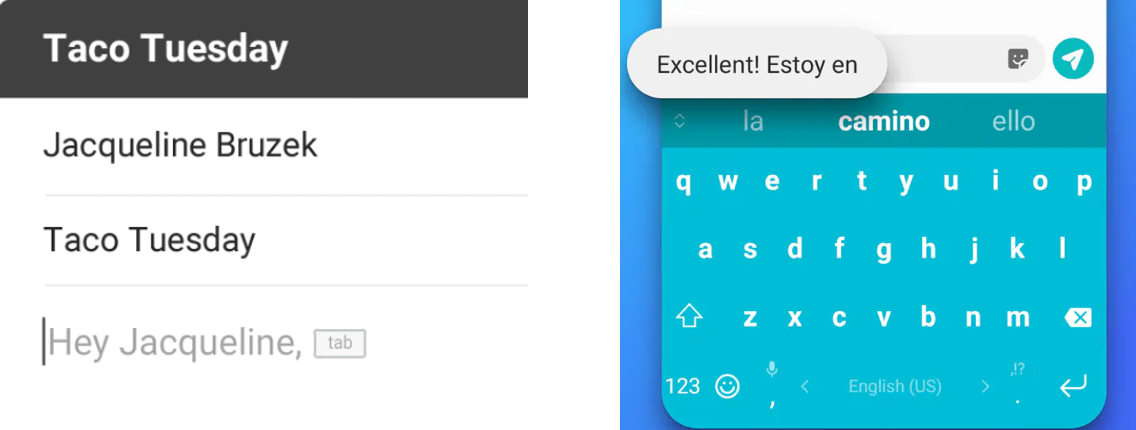}}
\caption{Two examples of language model deployments in practice. The figure on the left (image credit: \citep{fig_gmailsc}) is the Smart Compose feature for Gmail \citep{gmailsc} and the figure on the right (image credit: \citep{swiftkey}) is the Microsoft SwiftKey Keyboard.} 
\label{LM}
\end{center}
\vskip -0.2in
\end{figure}

Advances in language modeling have produced high-capacity models which perform very well on many language tasks. Language models are of particular interest as they are capable of generating free-form text, given a context, or even unprompted. There is a plethora of applications where language models have the opportunity to improve user experience, and many of them have been deployed in practice to do so, such as text auto-completion in emails and predictive keyboards (illustrated in Fig. \ref{LM}). Interestingly, language models with massive capacities have been shown to achieve strong performance in other tasks as well, e.g. translation, question-answering etc. even in a zero shot setting without fine-tuning in some cases \citep{brown2020language}.

On the other hand, recent studies have demonstrated that these models can memorize training samples, which can be subsequently reconstructed using probing attacks, or even during free-form generation \citep{secretsharer, carlini2020extracting}. While domain adaptation of general phrases is intended, the model should not leak or memorize rare sequences which could lead to a privacy breach according to GDPR, such as singling out of a user \citep{GDPR}.

Efforts to mitigate the risk that a model may yield rare samples which violate privacy include applying differential privacy (DP) during training \citep{dwork2011differential, song13, abadi2016deep}, as well as API hardening to ensure that attackers have little or no access to the model's underlying probability distributions.  While these approaches can be successful, it is challenging to quantify the residual privacy risks in language models, whether or not mitigations have been applied. In this work we propose a methodology for privacy investigations of a language model trained on confidential user content. Furthermore, we aim to produce metrics quantifying the likelihood a model might leak rare training data, under the strictest black-box assumptions about access to the model, i.e.\ that attackers can access only the model's top-$k$ prediction at each token position, given an input prefix. This choice of threat model enables us to assess a model's risk for realistic deployment scenarios, assuming best practices in API hardening are employed.  

\subsection{Contributions}
This paper makes the following contributions:
\begin{enumerate}
    \item We propose a methodology that investigates the user content in the training data that could be leaked by the model when prompted with the associated context in terms of user-level privacy.
    \item We introduce metrics that allow comparing models of various kinds (e.g. a DP model vs. a non-DP model) that are trained on the same training data in terms of privacy. 
    \item We demonstrate experimental results for both RNN and Transformer based models illustrating the application of our approach. We show how the proposed privacy investigation can provide valuable information towards protecting user-level privacy. We study the effects of mitigation techniques such as DP and API hardening through our metrics.
\end{enumerate}

The outline of the paper is as follows. In Section \ref{sec:background}, we provide background information for the language models focused in this work. Section \ref{sec:threat} defines the threat model and discusses the ability of an adversary towards attacking a language model deployed in practice. In Section \ref{sec:leakage_report}, we introduce our methodology of investigating a model trained on user content for the purpose of user-level privacy protection. We discuss special cases in Section \ref{sec:filter} and propose metrics to quantify user-level privacy leakage in Section \ref{sec:metrics}. We demonstrate our framework through numerical studies on real-world datasets in Section \ref{sec:tab_attack}. Section \ref{sec:related} discusses the related work and future directions and concludes the paper.

\section{Background: Language Models}
\label{sec:background}
Language modeling is the task of learning the underlying probability distribution over sequences of words in a natural language. A statistical model for a sequence of tokens $w_1, \ldots, w_n$ is represented by the joint probability $\Pr(w_1, \ldots, w_n)$, which can be further decomposed as the product of conditional probabilities:
\begin{align}
\label{factor}
    \Pr(w_1, \ldots, w_n) = \prod_{i=1}^{n} \Pr(w_i | w_1, \ldots, w_{i-1}).
\end{align}
Here $\Pr(w_i | w_1, \ldots, w_{i-1})$ represents the probability of the occurrence of token $w_i$ given the previous token sequence $w_1, \ldots, w_{i-1}$.

It has been shown that neural networks can be utilized to estimate these conditional distributions effectively and be employed as language models \citep{Bengio2003ANP}. Given an unsupervised corpus of tokens $\mathcal{W} = \{w_1, \ldots, w_n\}$, a standard language modeling objective is to maximize the following likelihood function:\footnote{The decomposition in \eqref{factor} is called forward autoregressive factorization. Although not all language models use this factorization in their training, we focus on the operation of the ones that could be deployed in practice for the generative text prediction task.}
\begin{align*}
 \mathcal{L}(\theta) = \sum_{i=1}^{n} \log \Pr \limits(w_i | w_1, \ldots, w_{i-1} ; \theta),
\end{align*}
where the conditional probability on $w_i$ is calculated by evaluating the neural network with parameters $\theta$ on the sequence $w_1, \ldots, w_{i-1}$.

The quality of a language model is commonly measured by two metrics, namely perplexity and top-$k$ accuracy. Perplexity measures the likelihood of text sequences and is defined as $\textnormal{PP}(w_1, \ldots, w_n) = 2^{-l}$ where
\begin{align*}
l = \frac{1}{n} \sum_{i=1}^{n} \log_2 \Pr \limits(w_i | w_1, \ldots, w_{i-1}).
\end{align*}
The evaluation of the perplexity on unseen data indicates how well the model fits the underlying distribution of the language. The smaller the value of perplexity, the better the language model is at modeling the data. Top-$k$ accuracy metric is defined as the ratio of the number of correct predictions to the total number of tokens\footnote{When $k>1$, correct prediction refers to the label being in the list of $k$ predictions returned by the model.}. The relevance of the parameter $k$ depends on the application. For instance, the accuracy for the highest-likelihood candidate (top-1 accuracy) is important for text auto-completion feature in emails \citep{gmailsc} whereas top-3 accuracy is also of interest for predictive keyboards (\citeauthor{swiftkey,gboard}).

There are a vast number of architectures employed for language models. At a high level, these architectures are either derived from variants of recurrent neural networks (RNNs) \citep{mikolov10,Sundermeyer_lstmneural,peters2018contextualized} or based on self-attention mechanisms of the transformer \citep{transformer,Radford2018ImprovingLU,howard-ruder-2018-universal,devlin-2019-bert,Yang2019,radford2019language,sun2019ernie,brown2020language,turingNLG}. Recently, large transformer based models have been achieving impressive state-of-the-art results in a variety of tasks \citep{brown2020language}. On the other hand, RNN based architectures might be favored in practice as well, e.g. when there are strict latency or memory requirements \citep{gmailsc}. 

\section{Threat Model}
\label{sec:threat}
Our threat model is tailored for privacy considerations when a language model is trained on confidential user content, which contains sensitive information that would lead to privacy violations in case they are leaked by the model \citep{GDPR,OSTP}. Such privacy considerations are in fact legitimate as language models perform next token prediction so they could be used in a generative fashion by entering a particular text prefix and asking the model to auto-complete indefinitely. Here, the danger is imminent as it is not \emph{a priori} clear what will be leaked from the user content in the training data. We note that any language model with non-zero utility will necessarily have the top-1 accuracy in the training data bounded away from zero. Therefore, ``something" will be leaked from the user content in the training data\footnote{We will use the terms ``leak" and ``correct prediction on the training data" for a model interchangeably in what follows.}.
Since the main objective of training language models is modeling the underlying distribution of a language, well-generalized models are not expected to memorize the rare sensitive information in the training data, as they are out-of-distribution and irrelevant to the learning task, hence unnecessary to improve the model performance. Recent results show that this is not the case \citep{secretsharer,carlini2020extracting,feldman2020,brown2020memorization,petroni2019}. When the data distribution is long-tailed (as is the natural language \citep{power_laws}), it has been shown that label memorization is necessary for achieving near-optimal accuracy on test data \citep{feldman2020,brown2020memorization}. Therefore, it is imperative to build privacy monitoring techniques to minimize the chances of an ``accidental" data leakage to prevent privacy violations.

\begin{figure}[ht]
\vskip 0.2in
\begin{center}
\centerline{\includegraphics[width=.85\columnwidth]{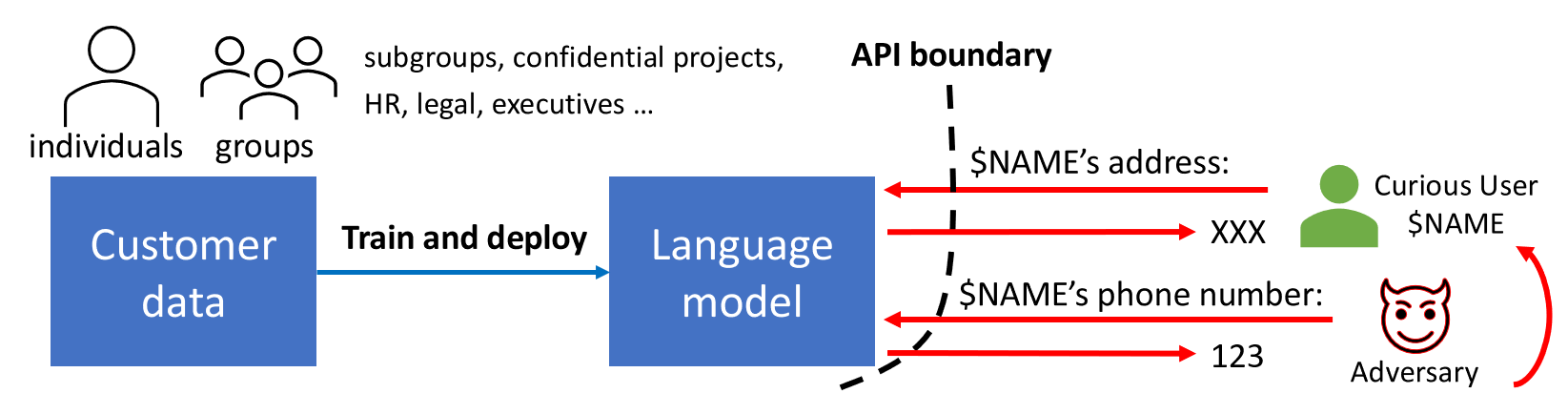}}
\caption{Illustration of our threat model. A language model is deployed after being trained on user content. One can query the model with a sequence of tokens and receive the top-$k$ predictions for the next token (top-1 in this example). A curious user (\$NAME) queries the model to see if their address is leaked by the model. More dangerously, an adversary inputs a directed query to learn the phone number of the targeted user.}
\label{fig:threat}
\end{center}
\vskip -0.2in
\end{figure}

Based on the discussion above, we consider a practical threat model that is relevant to the language models deployed in practice. We assume a black-box access, where a curious or malevolent user can query a pre-trained and deployed language model on any sequence of tokens $w_1, \ldots, w_{i}$ and receive the top-$k$ predictions returned by the model for the next token $w_{i+1}$. See Fig. \ref{fig:threat} for an illustrating example. We place no assumption on the parameter $k$ and let it be completely determined by the particular application for which the query is made (see Fig. \ref{LM}). We note that the threat model does not assume the availability of confidence scores or probabilities for the predictions and it is trivially applicable to the deployed models in practice. In fact, even the availability of the next token prediction(s) may not always be the case if the model does not return any prediction under certain conditions (e.g. when the prediction score is below a pre-fixed triggering threshold \citep{gmailsc}). However, since there is no guarantee that all sensitive information will be on the safe side of the triggering threshold, we believe it might be better to have protection against the worst case where the model prediction is available for the next token $w_{i+1}$ when any sequence of tokens $w_1, \ldots, w_{i}$ is queried.   

The threat model allows a curious user to know whether any sensitive information in their data is leaked by the model. Therefore, the data owner can use any prefix in their data to query the model. The threat model also includes the case of a malevolent user, who can input directed queries in order to extract sensitive information about a targeted user. 

\section{Training Data Leakage Analysis}
\label{sec:leakage_report}
\begin{figure}[t]
\vskip 0.2in
\begin{center}
\centerline{\includegraphics[width=.8\columnwidth]{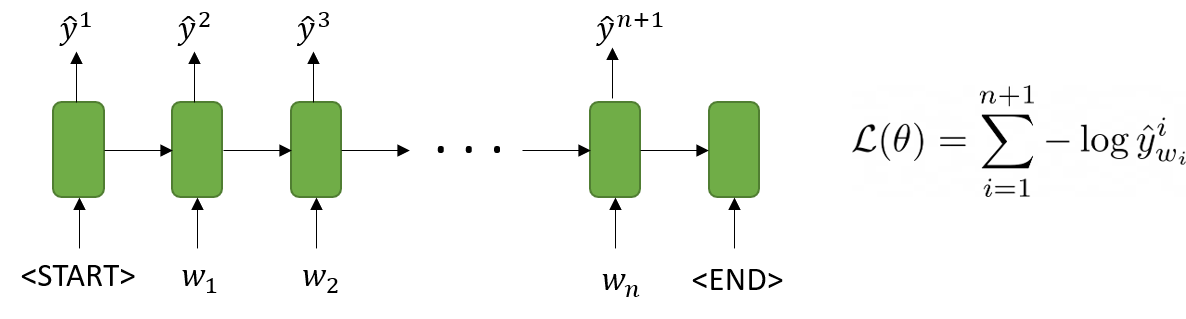}}
\caption{A typical way of training a language model on a sequence of tokens $w_1, \ldots, w_n$ (RNN type architecture is depicted for the sake of illustration). The model is yielding a probability distribution $\hat{y}^{i+1}$ having seen the context $w_1, \ldots, w_{i}$ to predict the next token $w_{i+1}$ for $i \in \{1, \ldots, n\}$. We note that the loss function is composed of loss at each time step, therefore, the model learns to predict $w_{i+1}$ with the context $w_1, \ldots, w_i$ for all $i \in \{1, \ldots, n\}$.} 
\label{fig:train}
\end{center}
\vskip -0.2in
\end{figure}

In this section, we introduce our framework to investigate a model trained on user content for the purpose of user-level privacy protection. We fix the notation first.
\paragraph{Notation} 
For a language model trained on user content, let $\mathcal{U} = \{U_1, \ldots, U_{n}\}$ specify the set of users. For each user $U_i \in \mathcal{U}$ we define the set $\mathcal{D}_i = \{ D_i^1, D_i^2, \ldots, D_i^{|\mathcal{D}_i|} \}$ ($|\mathcal{D}_i|$ refers to the size of the set $\mathcal{D}_i$) as the corresponding content on which the language model is trained. Each content $D_i^j$ is basically a sequence of tokens $w_1, w_2, \ldots, w_{|D_i^j|}$\footnote{Based on the preprocessing of the training data, this could be a sentence, a paragraph, or an email etc. We emphasize that $\mathcal{D}_i$ may even be considered as one large text sequence since it belongs to a single user and that is all it matters for our purpose.}. See Fig. \ref{fig:train} for an illustration of how a language model is typically trained on a sequence of tokens.  The training data $\mathcal{D}$ is the combination of all user content, i.e., $\mathcal{D} = \cup_{i \in \{1, 2, \ldots, n\}} \ \mathcal{D}_i = \cup_{i \in \{1, 2, \ldots, n\}, \ j \in \{1, 2, \ldots, |\mathcal{D}_i|\}} \ D_i^j$.

\begin{figure}[ht]
\vskip 0.2in
\begin{center}
\centerline{\includegraphics[width=.8\columnwidth]{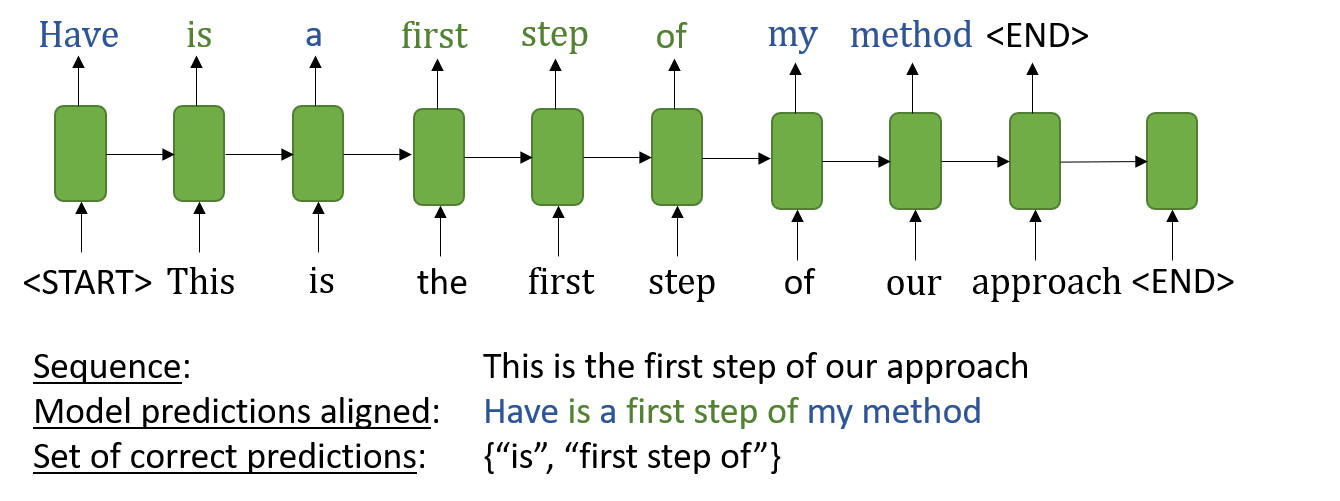}}
\caption{An illustration of the collection of correct model predictions. We run the model through each sequence in the training data and obtain the top-$k$ (top-1 in this example) prediction(s). We then collect the sequence of tokens where the model consecutively provides the correct prediction.} 
\label{fig:step1}
\end{center}
\vskip -0.2in
\end{figure}

We introduce our training data leakage analysis on a language model after being trained on the training data $\mathcal{D}$. The first step of our framework is to run the model through the training data and collect its correct predictions in the training data. We illustrate this step with an example in Fig. \ref{fig:step1}. This collection consists of sequences of tokens where the correct prediction is observed in top-$k$ predictions of the model consecutively. We emphasize that consecutive correct predictions is an important phenomenon because the longer the model leaks a training sequence $w_{i+1}, w_{i+2}, \ldots$ having seen the context $w_1, \ldots, w_i$, the more it discloses user content, causing privacy concerns. Therefore, we do not break sequences where the model provides correct predictions consecutively and collect all such sequences in the training data. In Algorithm \ref{alg:step1} we provide the pseudo-code to collect the correct predictions of the model as described above. Let us denote this collection as $\mathcal{S}$. We note that $\mathcal{S}$ is a multiset because it can contain multiple instances of a correctly predicted sequence in the training data\footnote{Henceforth we will use the terms set and multiset interchangeably for $\mathcal{S}$.}.

The main part of our training data leakage analysis is to provide key features for each sequence in the set $\mathcal{S}$, which contain important information for privacy investigations. 
We next describe each feature in detail and provide the accompanying Table \ref{table:report} as an illustrative example. 

\begin{table}[b]
\caption{An artificial example to describe the features of our training data leakage analysis. In this example, there is a sequence ``very much" appearing two times in $\mathcal{S}$, meaning that it is predicted correctly two times by the model. These correct predictions appear in a single user's data (either in a single $D_i^j$ or two different $D_i^{j_1}$ and $D_i^{j_2}$ for some user $U_i$). The corresponding contexts on which the model produces this correct sequence are ``Thank you" and ``I like cats" and the corresponding perplexities are 1.3 and 3.6. On the other hand, the sequence ``very much" itself appears ten times in the training data $\mathcal{D}$ (only in two of which the model predicts the sequence correctly), among five user's data.}
\label{table:report}
\vskip 0.15in
\begin{center}
\begin{small}
\begin{sc}
\begin{tabular}{p{0.12\textwidth}p{0.11\textwidth}p{0.1\textwidth}p{0.11\textwidth}p{0.1\textwidth}p{0.14\textwidth}p{0.08\textwidth}}
\toprule
$\mathcal{S}$ & total \# in $\mathcal{S}$ & user \# in $\mathcal{S}$ & total \# in $\mathcal{D}$ & user \# in $\mathcal{D}$ & context(s) & perp. \\
\midrule
``very much" & 2 & 1 & 10 & 5 & [``Thank you", ``I like cats"] & [1.3, 3.6] \\
\bottomrule
\end{tabular}
\end{sc}
\end{small}
\end{center}
\vskip -0.1in
\end{table}

\paragraph{total count in $\mathcal{S}$:} This is simply the number of occurrences of a sequence in the multiset $\mathcal{S}$. This feature shows in total how many times the model leaks a sequence given the correct context. A large number may seem to indicate that the sequence is common and not concerning for privacy, however, that may not be the case if it is present in only a single user's data, which is captured in the next feature.
\paragraph{user count in $\mathcal{S}$:} For any sequence in $\mathcal{S}$ where the total count is larger than one, here we count the number of distinct users for which \textit{the correct prediction of this sequence is made}. This is important because a sensitive information of a single user may appear multiple times in their data (e.g. address being emailed a number of times), which can be memorized by the model. Along with the previous feature we can investigate such cases effectively with this feature. 
\paragraph{total count in $\mathcal{D}$:} In the previous two features, we count the occurrences in the set $\mathcal{S}$. In the next two features, we count the occurrences in the training data $\mathcal{D}$. This may be helpful because for a sequence that is predicted correctly only for a single user, i.e. user count in $\mathcal{S}$ is one, this may not immediately imply that a sensitive information is leaked by the model. It may be the case that the sequence appears many times among various users' data, which could provide ``plausible deniability" in the sense that many other users contributed the model to learn this sequence. This can be calculated by simple string matching and formally expressed as $\sum \limits_{i=1}^{n} \sum \limits_{j=1}^{|\mathcal{D}_i|} (\textnormal{count of $w$ in $D_i^j$})$ for a sequence $w \in \mathcal{S}$. 
\paragraph{user count in $\mathcal{D}$:} Connected to the previous feature, here we count the number of distinct users for which a sequence in $\mathcal{S}$ is found in their data. We note that here we do not consider whether the model correctly predicts the sequence or not given the right context, that was done in user count in $\mathcal{S}$. Instead, we calculate this by simply checking if the sequence can be found in a user's data via string match. This can be formally expressed as $\sum \limits_{i=1}^{n} 1(\textnormal{$\exists j \in \{1, 2, \ldots, |\mathcal{D}_i|\}$ s.t. $w \subseteq D_i^j$})$ where $1(\cdot)$ is an indicator function for a sequence $w \in \mathcal{S}$. Sequences for which the user count in $\mathcal{D}$ is large are unlikely to be sensitive for a single user. However, there might still be concerning cases, e.g., the model predicts the sequence correctly only for a single user (i.e. user count in $\mathcal{S}$ is one) although there is plausible deniability as discussed previously.
\paragraph{context(s)} For any sequence in $\mathcal{S}$, this feature provides the corresponding context(s) with which when prompted the model it predicts the sequence correctly. This feature is useful to check for instance if long sequences can be found with short contexts, which means that the model leaks a long user content when prompted with a short initial context.
\paragraph{perplexity(ies)} Connected to the feature above, this feature provides the corresponding perplexity(ies) for any sequence in $\mathcal{S}$. This is also an important feature as it shows how certain the model is when predicting a sequence. Furthermore, it allows comparing the correct predictions of the model with a public model\footnote{Public model refers to a language model trained on a public dataset.}. Considering a sequence in $\mathcal{S}$ where the user count in $\mathcal{D}$ is one, a small perplexity on the language model along with a large perplexity on public model might indicate that a sensitive information is leaked about the corresponding user since the sequence is ``surprising" to the public model by the large perplexity.

\begin{algorithm}[b]
   \caption{The collection of correct model predictions.}
   \label{alg:step1}
\begin{algorithmic}
   \STATE {\bfseries Input:} A language model $LM(\cdot)$ and the corresponding training data $\mathcal{D}$
   \STATE {\bfseries Output:} The (multi)set $\mathcal{S}$ of correct predictions
   \STATE Initialize $\mathcal{S} = []$
   \FOR{$i=1$ {\bfseries to} $n$}
   \FOR{$j=1$ {\bfseries to} $|\mathcal{D}_i|$}
   \STATE Initialize $W = ``"$
   \STATE Let $D_i^j = [w_1, \ldots, w_{|D_i^j|}]$
   \FOR{$l=1$ {\bfseries to} $|D_i^j|$}
   \STATE Obtain top-$k$ predictions $preds = LM(D_i^j[:l])$
   \IF{$w_l \in preds$}
   \STATE Append $w_l$ to $W$
   \ELSIF{$W \neq ``"$}
   \STATE Append $W$ to $\mathcal{S}$ and initialize $W = ``"$
   \ENDIF
   \ENDFOR
   \ENDFOR
   \ENDFOR
\end{algorithmic}
\end{algorithm}

Before concluding this section, we discuss a number of points regarding our training data leakage analysis. 
\begin{itemize}
    \item In our framework, the sequences in $\mathcal{S}$ are created when the model is prompted with the ``right" context, i.e. the context on which the model is trained for next token prediction task. However, a sensitive information might even be leaked under a context not appearing in the training data. Such a case might be missed if the same sensitive information is predicted wrong under its corresponding context in the training data, therefore not being included in $\mathcal{S}$. Looking at the loss function in Fig. \ref{fig:train}, intuitively one can expect that the model would more likely predict the sequence under the context it has seen during training than any other context, however, there is also no guarantee that this will always be the case. Therefore, it might be worthwhile to extend the set of contexts beyond the training data and then create the set $\mathcal{S}$. This is an interesting future direction that could strengthen the privacy investigation of a language model.
    \item Such a detailed investigation of sequences in our training data leakage analysis may not be possible if the model training is done with no access to look at the training data \citep{gmailsc}. In that case, we can simply replace each sequence and its corresponding context(s) with their length of tokens and still obtain valuable information. For example, we can investigate the length of the sequences for which the user count in $\mathcal{S}$ is small to see if long completions are possible or check the length of the contexts to see if the leakage is possible with short contexts. The perplexities could also be very helpful in this case because we can measure how surprising each sequence is to a public model that is only trained on public dataset(s). If the perplexities are similar, then this indicates that the prediction is as ``familiar" to the public model, therefore, unlikely to be a privacy concern for a user.
    \item While Algorithm \ref{alg:step1} sequentially goes over the training data to illustrate our approach, we point out that all the steps from the collection of correct model predictions to the counts performed over the training set are fully parallelizable for an efficient implementation. We provide a more detailed discussion on the complexity of our approach in Appendix \ref{sec:complexity}.
\end{itemize}


\section{Filtering for Special Cases}
\label{sec:filter}
Our leakage analysis may include a massive number of sequences when the training data $\mathcal{D}$ is large, making privacy investigations infeasible. However, simple filtering procedures can be applied to reduce the number of sequences effectively. For instance, keeping the ones for which the user count in $\mathcal{S}$ is less than $p$ for some threshold $p$ of interest is a reasonable operation. The sequences filtered here would the ones where there are at least $p$ users having this sequence in their data and the model predicts the sequence correctly for each of these users. From the perspective of model predictions, this is reminiscent of $k$-anonymity \citep{sweeney2002k} as a famous data anonymization technique\footnote{Since the contexts will likely be different, $k$-anonymity is still substantially more powerful for anonymization.}. 

Another important case is regarding the sequences in $\mathcal{S}$ for which the user count in $\mathcal{D}$ is one, i.e., there is only one user having this sequence in their data and the model leaks the sequence when prompted with the corresponding context\footnote{or contexts. Note that the total count in $\mathcal{D}$ and total count in $\mathcal{S}$ can still be arbitrary since the sequence can appear multiple times in the user's data. Nevertheless, since it belongs to only one user, the sequence should be protected equally.}. This is inarguably the case with the most potential to result in privacy violations since leakage of a unique content of a user may lead to singling out of that user. Let us denote the set of sequences in $\mathcal{S}$ that are unique to a user as $\mathcal{S}_{\textnormal{uniq}}$. In the next section, we focus on the set $\mathcal{S}_{\textnormal{uniq}}$ and propose metrics that quantify user-level privacy leakage.

\section{Metrics to Quantify Privacy Leakage}
\label{sec:metrics}
In this section, we propose two metrics to quantify user-level privacy leakage, which are straightforward to interpret, and compare different models trained on the same training data in terms of privacy:
\begin{enumerate}
    \item The first metric is the number of sequences in  $\mathcal{S}_{\textnormal{uniq}}$, i.e. the sequences in $\mathcal{S}$ for which the user count in $\mathcal{D}$ is one.
    \item The second metric is a curated version of the first metric. We still consider the sequences in the set $\mathcal{S}_{\textnormal{uniq}}$ but we remove the ones for which the ratio of the perplexity with respect to a public model and our language model is below some threshold $t$, i.e. ${\textnormal{PP}_{\textnormal{public}}(w)}/{\textnormal{PP}_{\textnormal{lm}}(w)} < t$. This basically filters out the unique sequences that have similar perplexities with respect to a public model as there is plausible deniability of similar leakage possibility, given a public model. We further define the worst-case \emph{leakage epsilon}
    \begin{align}
    \epsilon_l \triangleq \max \limits_{w \in \mathcal{S}_{\textnormal{uniq}}} \log \left( \dfrac{\textnormal{PP}_{\textnormal{public}}(w)}{\textnormal{PP}_{\textnormal{lm}}(w)} \right), 
    \label{leakage_epsilon}    
    \end{align}
    measuring the perplexity ratio with respect to a public model maximized over the sequences in the set $\mathcal{S}_{\textnormal{uniq}}$ to capture the worst-case scenario. 
\end{enumerate}

We next discuss the advantages and disadvantages of the proposed metrics. The first metric is simple and easy to use. However, we observe in our experiments that even a public model that is not trained on a private data can predict unique sequences in the private data. Such unique sequences would likely not constitute a privacy violation since the public model has not seen any private data in its training. The disadvantage of this metric is that it does not consider the likelihood that sequences in $\mathcal{S}_{\textnormal{uniq}}$ might also be easily predicted by a public model, reducing potential privacy risks.


Our second metric touches on this point by eliminating the sequences that do not look surprising to a public model. However, the main disadvantage of this metric is the hardship of the choice of the threshold $t$. Therefore, we introduce the term leakage epsilon, denoted by $\epsilon_l$, measuring the worst-case perplexity ratio with respect to a public model over the unique sequences. This is motivated by the definition of differential privacy (DP), which bounds the worst-case effect of a single substitution in the data. DP is a strong framework providing global guarantees for an algorithm over all possible users and all possible training sets. On the other hand, Eq. \eqref{leakage_epsilon} compares the relative likelihoods of model outputs with respect to a public reference model and introduces a metric to compare different models (including non-DP ones) in terms of privacy. A smaller $\epsilon_l$ for a language model translates into a better privacy protection as the unique sequences leaked by the model will have relatively similar perplexities with respect to a public model, providing plausible deniability for each one of them.

The final point related to both metrics is the choice of the public model. We note that $\epsilon_l$ depends on the public model used in Eq. \eqref{leakage_epsilon}. If the distribution of the public dataset differs substantially from that of the private one used to train the language model, this may lead to a pessimistic $\epsilon_l$ value. To circumvent this, one might use a collection of public models ($\mathcal{P}$) trained on various public datasets and modify the term as $\epsilon_l \triangleq \max \limits_{w \in \mathcal{S}_{\textnormal{uniq}}} \min \limits_{\textnormal{public} \in \mathcal{P}} \log \left( {\textnormal{PP}_{\textnormal{public}}(w)}/{\textnormal{PP}_{\textnormal{lm}}(w)} \right)$. One can further consider removing the users for which the language model leaks a unique sequence (i.e. the users in $\mathcal{S}_{\textnormal{uniq}}$) and all their data from the training data and train another model on the remaining users. This latter model may be employed as a ``public model" in Eq. \eqref{leakage_epsilon} to calculate $\epsilon_l$ as it has not seen any data of any user in $\mathcal{S}_{\textnormal{uniq}}$ during its training. This is also motivated from differential privacy, which provides the guarantee that removing a user from the dataset will not change the probability distribution of the output of an algorithm substantially (bounded by $\exp(\epsilon)$). 

\section{Case Study: Tab Attack}
\label{sec:tab_attack}
In this section, we provide case studies of our leakage analysis. We consider an attack setting that has access to top-1 predictions of a language model. Having in mind the text auto-completion feature in emails where the predictions are applied by pressing the TAB key on the keyboard (see Fig. \ref{LM}), we dub this as the \emph{tab attack}. We investigate the unique sequences ($\mathcal{S}_{\textnormal{uniq}}$) that could be leaked via the tab attack when the model is queried with the corresponding context. We note that although the attacker ability is limited to top-1 predictions, the model builder can utilize all information to investigate the unique sequences that could be leaked by the tab attack prior to the model deployment. We apply our leakage analysis to the unique sequences ($\mathcal{S}_{\textnormal{uniq}}$) to assess the attack surface under the tab attack threat model.


\subsection{RNN-based model on Reddit dataset}
\label{sec:reddit}
We study a large-scale example as a realistic setup for the deployed language models in practice.

\paragraph{Dataset} We use a large dataset of Reddit posts, as decribed by \citet{alrfou2016conversational}, that contains 140M sentences from 4.4M users for a randomly chosen month (Oct 2018). It is randomly split into 90\% training and 10\% validation sets. Reddit dataset is commonly used in privacy research since each post in the dataset is keyed by a user so the data can be grouped by users to provide user-level privacy.
We provide three sets of language models trained on this private Reddit dataset.
\begin{enumerate}
    \item A language model trained on the Reddit dataset. This will be referred to as \textit{Private LM} in our results.
    \item A language model trained on the Reddit dataset with differential privacy \citep{dprnn, abadi2016deep}. We take four snapshots of the model during training, corresponding to four DP language models with epsilons 3.28, 4.68, 6.20, and 26.4\footnote{The models satisfy user-level DP and $\delta \lesssim 1/(\textnormal{\# users})$ same for all models.}. The training begins with a random initialization of the weights. The models will be referred to as \textit{DP-LM RanIni $\epsilon = \cdot$}.
    \item A language model trained on the Reddit dataset with differential privacy. Here, the model weights are initialized from a public model trained on Google News dataset \citep{billionword}. It has been shown that transfer learning helps obtaining strong privacy guarantees with a minor cost in utility \citep{dprnn, abadi2016deep, tramer2020differentially, papernot2020}. We similarly take three snapshots of the model during training, corresponding to three DP language models with epsilons 2.98, 4.47, and 6.68. These models and the public model will be referred to as \textit{DP-LM PubIni $\epsilon = \cdot$} and \textit{Public LM} respectively.
\end{enumerate}

\begin{table}[ht]
\caption{Results of the experiment on RNN-based language models trained on Reddit dataset \citep{alrfou2016conversational}. We provide the perplexity and accuracy on the validation set to compare the performances of the models. In the next column, we provide the number of unique sequences (i.e. $|\mathcal{S}_{\textnormal{uniq}}|$) for each model. We calculate leakage epsilon $\epsilon_l$ for some of the models for comparison in the last column.}
\label{table:reddit}
\vskip 0.15in
\begin{center}
\begin{small}
\begin{sc}
\begin{tabular}{p{0.3\textwidth}p{0.12\textwidth}p{0.12\textwidth}p{0.2\textwidth}p{0.05\textwidth}}
\toprule
Model & Val perp & Val acc (\%) & \# unique seq. $\left(|\mathcal{S}_{\textnormal{uniq}}|\right)$ & $\epsilon_l$ \\
\midrule
Private LM & 69.4 & 23.7 & 3757 & 17.75 \\
DP-LM RanIni $\epsilon=3.28$ & 290.0 & 14.5 & 0 & - \\
DP-LM RanIni $\epsilon=4.68$ & 130.3 & 19.6 & 5 & - \\
DP-LM RanIni $\epsilon=6.20$ & 107.8 & 20.8 & 11 & 0.64 \\
DP-LM RanIni $\epsilon=26.40$ & 96.5 & 21.5 & 30 & - \\
Public LM & 757.5 & 13.1 & 159 & - \\
DP-LM PubIni $\epsilon=2.98$ & 183.1 & 19.7 & 157 & 0.29 \\
DP-LM PubIni $\epsilon=4.47$ & 106.7 & 21.9 & 203 & - \\
DP-LM PubIni $\epsilon=6.68$ & 92.8 & 22.2 & 246 & 1.33 \\
\bottomrule
\end{tabular}
\end{sc}
\end{small}
\end{center}
\vskip -0.1in
\end{table}

The model architecture is same for all these models and the details are specified below. 

\paragraph{Model} We use a one-layer GRU model as the language model for the next-word prediction task. The embedding size is set to 160 and the hidden size to 512, and the vocabulary is fixed to the most frequent 10k words in the training corpus (out of 3.2M words). We use the Adamax optimizer with the learning rate set to 1e-3 and the batch size is set to 3072 in the DP training and to 512 otherwise.

We provide in Table \ref{table:reddit} the performances of the models and the result of the tab attack for each of them. We discuss the results of this experiment in what follows.

We observe from Table \ref{table:reddit} that the private LM that is trained without DP leaks a huge number of unique sequences (3757) from the training data. There are 759 unique sequences for which the number of tokens is larger than 9. A majority of these examples are coming from highly-repeated sentences (728 of these sequences are repeated somewhere between 50-34372 times) by the bots in the Reddit dataset\footnote{An example of a unique sequence memorized by the model is ``has been automatically removed because the title does not include one of the required tags ." repeated 5377 times in the bot's data.}. Expectedly, the resulting $\epsilon_l = 17.75$ is very large and it does not offer any reasonable privacy guarantee. We calculated $\epsilon_l$ among the unique sequences that do not repeat more than once and found that it is $4.60$. This shows the importance of de-duplication at a granular level (e.g. removal of sentence duplicates) as also observed by \citet{secretsharer,carlini2020extracting}.

For the DP-LMs that are snapshots of a model trained with random initialization of weights, we observe small number of unique sequences leaked by the models. Interestingly, we get no unique sequence with the first one having $\epsilon = 3.28$, although there is a high cost in terms of utility. We provide the list of unique sequences for the models with $\epsilon = 4.68, 6.20$ and $26.4$ in Table \ref{table:eps_4_68}, \ref{table:eps_6_20}, and \ref{table:eps_26_4} of Appendix \ref{sec:a2} respectively. We observe the efficacy of user-level DP training by noting that the unique sequences with large repetitions that were memorized by the private model have all disappeared with DP-LMs. Furthermore, there is a substantial decrease in the number of unique sequences, even for the DP-LM with relatively high epsilon value $\epsilon = 26.4$, which does not provide a reasonable theoretical privacy guarantee.

A interesting phenomenon we have observed consistently over all experiments is about the punctuation. 
Almost all unique sequences for the DP-LMs presented in Appendix \ref{sec:a2} have a punctuation mark. In our experiments we did not exclude the punctuation from the model predictions and treated them as any other token in the training data.

For the DP-LMs initialized from a public model, we observe relatively larger number of unique sequences leaked by the models. However, this is not surprising as the public model itself can predict 159 unique sequences in the private data, without seeing any private data in its training. Since the DP training is initialized from the public model in this case, it should be expected to get larger number of unique sequences. We note that leakage epsilon $\epsilon_l$ may provide a better ground for a fair comparison of models trained in different ways (e.g. random initialization vs. transfer learning).

We calculate leakage epsilon $\epsilon_l$ for three DP-LMs for comparison. The public models in the calculation of Eq. \eqref{leakage_epsilon} are as follows. For each model, we take the users who are the owners of the unique sequences leaked by the model and remove all their data from the training data. We subsequently train a new model on the remaining users. We consider the new model as the public model for the users of the unique sequences since it has not seen any data of these users during its training. We observe that DP-LM PubIni $\epsilon=2.98$ model has $\epsilon_l = 0.29$, much smaller than the models DP-LM RanIni $\epsilon=6.20$ with $\epsilon_l = 0.64$ and DP-LM PubIni $\epsilon=6.68$ with $\epsilon_l = 1.33$. This may not be surprising since $\epsilon=2.98$ provides much stronger privacy guarantees compared to $\epsilon=6.20$ and $\epsilon=6.68$. We note that all three models have quite small $\epsilon_l$ values, indicating that the unique sequences leaked by these models can also be simply learned from other users because they have similar perplexities with respect to the public model.

\subsection{Transformer-based model on Reddit dataset}

We next study a Transformer-based language model in a similar setting as the previous section.

\paragraph{Model} We use a GPT2 type model \citep{radford2019language} with six transformer layers with twelve attention heads each with a hidden dimension of 768 totalling about 82M parameters.
The data is tokenized using a byte-level version of Byte Pair Encoding (BPE) and a vocabulary size of 50257. The freely available base model,\footnote{\url{https://huggingface.co/distilgpt2}} which was trained on the OpenWebText and wikitext datasets \citep{Gokaslan2019OpenWeb, wikitext}, acts as the public model.

\paragraph{Dataset} We choose a random month (Jan 2019) of the Reddit dataset similar to the experiment in the previous section.
We drop posts which are shorter than 80 tokens and cap long posts to 256 tokens.
The final training dataset contains 1.5M posts from 1.3M users.
We perform a similar pre-processing step on the data for Feb 2019 and use 50,000 samples as the validation set.

We use three language models to perform a similar analysis. In addition to the public model, we fine-tune the public model with and without differential privacy on the private Reddit dataset and refer to them as DP-LM and Private LM respectively.
The DP language model provides good user-level privacy with $\epsilon = 1$ and $\delta = 10^{-7}$.

\begin{table}[ht]
\caption{Results of the experiment on Transformer-based language models. The columns follow similarly as in Table \ref{table:reddit}.}
\label{table:distilgpt2}
\vskip 0.15in
\begin{center}
\begin{small}
\begin{sc}
\begin{tabular}{p{0.2\textwidth}p{0.1\textwidth}p{0.21\textwidth}p{0.05\textwidth}}
\toprule
Model & Val perp  & \# unique seq. $\left(|\mathcal{S}_{\textnormal{uniq}}|\right)$ & $\epsilon_l$ \\
\midrule
Private LM & 49.7 & 333099 & 11.96 \\
DP-LM & 81.0 & 150247 & 2.25 \\
Public LM & 167.9& 115920 & - \\
\bottomrule
\end{tabular}
\end{sc}
\end{small}
\end{center}
\vskip -0.1in
\end{table}

We present in Table \ref{table:distilgpt2} the results of the experiment. We first note that even the public model produces an enormous number of unique sequences. This is because (i) BPE tokenizer has no out-of-vocabulary tokens, thus creating myriad unique sub-word token sequences in the training data and (ii) Transformer-based models have the attention mechanism, which gives a strong ability to copy tokens from context to predict the next token. This highlights the need for our metrics introduced in Section \ref{sec:metrics} to flag leaked unique sequences which can be considered not a privacy issue.

We observe that DP indeed offers a reliable way of reducing the leakage. We have selected a model of acceptable perplexity of 81 while maintaining good user-level epsilon of 1. The corresponding leakage epsilon is $\epsilon_l = 2.25$, significantly improving to that of the Private LM with $\epsilon_l = 11.96$ (recall that $\epsilon_l$ is log-based). 

Although there is no sentence-level repetition in the dataset, the reason that Private LM still has very high epsilon leakage may be attributed to the strong memorization capability of Transformer-based models. Furthermore, sub-word tokenization strongly necessitates personally identifiable information (PII) scrubbing of the training data, as from our observations many unique sequences leaked by Private LM for which the perplexity ratio in Eq. \eqref{leakage_epsilon} is large contain PII that can easily lead to singling out of a user. 



\section{Related Work and Conclusion}
\label{sec:related}
A wide body of work has demonstrated privacy issues in general for machine learning models trained on personal data. Language models are among the most to suffer as they are capable of generating text which may potentially leak sensitive user content and lead to serious privacy violations. 

\citet{zhang2016} show that deep learning models can achieve perfect accuracy even on randomly labeled data.
Such memorization capability may in fact be needed to achieve near-optimal accuracy on test data when the data distribution is long-tailed as recently shown by \citet{feldman2020,brown2020memorization}.
Unfortunately this can lead to a successful training data extraction attack, as in the case for the concurrent work \citep{carlini2020extracting} that can recover individual training examples from the GPT-2 language model \citep{radford2019language}.
In their method, \citet{carlini2020extracting} generate a list of sequences by sampling from the GPT-2 language model and then curate it by using the perplexity measure.
In a related line of work which exploits the increasingly common transfer learning setup, \citet{snapshotattack} have demonstrated that having simultaneous black box access to the pre-trained and fine-tuned language models allows them to extract rare sequences from the smaller and typically more sensitive fine-tuning dataset.
Both attacks rely on the model output beyond top-1 or top-3 predictions along with the perplexity measure.
Access to this information may easily be restricted in deployed language models.
Nevertheless, there are serious privacy concerns since the attacks can extract personally identifiable information even if they are present in one document in the training data.
We believe that our proposed procedure for privacy investigations of a language model trained on user content could be very beneficial to protect user-level privacy in the presence of such attacks.

On the other hand, \citet{secretsharer} introduced the exposure metric to quantitatively assess the unintentional memorization phenomenon occurring in generative sequence models. They do so by inserting randomly-chosen canary sequences a varying number of times into the training data and measuring the relative difference in perplexity between inserted canaries and non-inserted random sequences. Our work is complementary in the sense that we are investigating the information leaked from user content in the training data, having in mind a strong threat model where one can query the language model with the precise context appearing in the training data. We believe that our proposed metrics along with the exposure metric can be employed together to provide strong privacy guarantees for a deployed language model.  

Another line of work has studied the vulnerability of machine learning models to membership inference attack \citep{shokri17, yeom18,  song19, nasr19, long18, LOGAN, truex18, irolla19, hisamoto20, salem18, sablayrolles19a, Leino20, choo2020labelonly}. The goal is to determine if a particular data record (or more generally data of a given user) belongs to the training set of the model. Although being an indirect leakage, membership inference is considered as a confidentiality violation and potential threat to the training data from models \citep{murakonda2020ml}.

The main framework with theoretical guarantees for user-level privacy is the application of differential privacy (DP) \citep{dwork2011differential} to model training. DP makes provable guarantees about the privacy of a stochastic function of a given dataset. Differentially private stochastic gradient descent (DP-SGD) has been developed and applied to training machine learning models \citep{song13, abadi2016deep}. This is an active area of research with the goal of pushing the frontiers of privacy-utility trade-off for deep neural networks.

\subsection{Future work}

We discuss a number of interesting future directions following our work:
\begin{itemize}
\item The proposed leakage analysis is based on central learning setting where the training data is stored at a central server. It would be interesting to solve the challenge of applying this method to other settings, such as federated learning \citep{kairouz19} where machine learning models are trained on decentralized on-device data.
\item We are hopeful that the metrics proposed in this work, as a first attempt to quantify user-level privacy leakage, would initiate further research on the topic, which will lead to further improvements on these metrics. 
\item It would be valuable to study the proposed methodology on more models/datasets, which would shed new lights on the protection of user-level privacy when language models are trained on confidential user content.  
\end{itemize}

\subsection{Conclusion}

Recent results show that language models are capable of memorizing training samples under the hood of their impressive performance. This poses an immediate threat as leaking rare user content could lead to a privacy breach according to regulations such as GDPR, e.g. due to singling out of a user. 

This work introduced a methodology to investigate information leaked by a language model from its training data in terms of privacy. We proposed metrics that could be used to quantify user-level privacy leakage and allow comparing models trained on the same data in terms of privacy. We believe our framework can be incorporated into the training platform of language models that would help assess the model from the perspective of privacy, along with its utility.   

\bibliography{references}

\begin{thebibliography}{53}
\providecommand{\natexlab}[1]{#1}
\providecommand{\url}[1]{\texttt{#1}}
\expandafter\ifx\csname urlstyle\endcsname\relax
  \providecommand{\doi}[1]{doi: #1}\else
  \providecommand{\doi}{doi: \begingroup \urlstyle{rm}\Url}\fi

\bibitem[Brown et~al.(2020{\natexlab{a}})Brown, Mann, Ryder, and
  Subbiah]{brown2020language}
Tom~B. Brown, Benjamin Mann, Nick Ryder, and Melanie Subbiah.
\newblock Language models are few-shot learners.
\newblock \emph{arXiv preprint arXiv:2005.14165}, 2020{\natexlab{a}}.

\bibitem[Carlini et~al.(2019)Carlini, Liu, Erlingsson, Kos, and
  Song]{secretsharer}
Nicholas Carlini, Chang Liu, {\'U}lfar Erlingsson, Jernej Kos, and Dawn Song.
\newblock The {S}ecret {S}harer: Evaluating and testing unintended memorization
  in neural networks.
\newblock In \emph{{USENIX} Security 2019}, August 2019.

\bibitem[Carlini et~al.(2020)Carlini, Tramer, Wallace, Jagielski, Herbert-Voss,
  Lee, Roberts, Brown, Song, Erlingsson, Oprea, and
  Raffel]{carlini2020extracting}
Nicholas Carlini, Florian Tramer, Eric Wallace, Matthew Jagielski, Ariel
  Herbert-Voss, Katherine Lee, Adam Roberts, Tom Brown, Dawn Song, Ulfar
  Erlingsson, Alina Oprea, and Colin Raffel.
\newblock Extracting training data from large language models.
\newblock \emph{arXiv preprint arXiv:2012.07805}, 2020.

\bibitem[{Art. 29 WP}(2014)]{GDPR}
{Art. 29 WP}.
\newblock Opinion 05/2014 on ``{A}nonymisation {T}echniques'', 2014.
\newblock URL
  \url{https://ec.europa.eu/justice/article-29/documentation/opinion-recommendation/files/2014/wp216_en.pdf}.

\bibitem[Lambert(2018)]{fig_gmailsc}
Paul Lambert.
\newblock {S}{U}{B}{J}{E}{C}{T}: {W}rite emails faster with {S}mart {C}ompose
  in {G}mail, 2018.
\newblock URL
  \url{https://www.blog.google/products/gmail/subject-write-emails-faster-smart-compose-gmail}.

\bibitem[Chen et~al.(2019)Chen, Lee, Bansal, Cao, Zhang, Lu, Tsay, Wang, Dai,
  Chen, Sohn, and Wu]{gmailsc}
Mia~Xu Chen, Benjamin~N. Lee, Gagan Bansal, Yuan Cao, Shuyuan Zhang, Justin Lu,
  Jackie Tsay, Yinan Wang, Andrew~M. Dai, Zhifeng Chen, Timothy Sohn, and
  Yonghui Wu.
\newblock Gmail {S}mart {C}ompose: Real-time assisted writing.
\newblock KDD '19, page 2287–2295, New York, NY, USA, 2019.

\bibitem[{M}icrosoft {S}wiftKey()]{swiftkey}
{M}icrosoft {S}wiftKey.
\newblock URL \url{https://www.microsoft.com/en-us/swiftkey}.

\bibitem[Dwork(2011)]{dwork2011differential}
Cynthia Dwork.
\newblock Differential privacy.
\newblock \emph{Encyclopedia of Cryptography and Security}, 2011.

\bibitem[Song et~al.(2013)Song, Chaudhuri, and Sarwate]{song13}
Shuang Song, Kamalika Chaudhuri, and Anand~D. Sarwate.
\newblock Stochastic gradient descent with differentially private updates.
\newblock In \emph{2013 IEEE Global Conf. on Signal and Information
  Processing}, pages 245--248, 2013.

\bibitem[Abadi et~al.(2016)Abadi, Chu, Goodfellow, McMahan, Mironov, Talwar,
  and Zhang]{abadi2016deep}
Martin Abadi, Andy Chu, Ian Goodfellow, H~Brendan McMahan, Ilya Mironov, Kunal
  Talwar, and Li~Zhang.
\newblock Deep learning with differential privacy.
\newblock In \emph{ACM CCS}, 2016.

\bibitem[Bengio et~al.(2003)Bengio, Ducharme, Vincent, and
  Jauvin]{Bengio2003ANP}
Yoshua Bengio, R{\'e}jean Ducharme, Pascal Vincent, and Christian Jauvin.
\newblock A neural probabilistic language model.
\newblock \emph{Journal of machine learning research}, 3:\penalty0 1137--1155,
  2003.

\bibitem[{G}board()]{gboard}
{G}board.
\newblock {T}he {M}achine {I}ntelligence {B}ehind {G}board.
\newblock URL
  \url{https://ai.googleblog.com/2017/05/the-machine-intelligence-behind-gboard.html}.

\bibitem[Mikolov et~al.(2010)Mikolov, Karafiát, Burget, Černocký, and
  Khudanpur]{mikolov10}
Tomáš Mikolov, Martin Karafiát, Lukáš Burget, Jan~``Honza" Černocký, and
  Sanjeev Khudanpur.
\newblock Recurrent neural network based language model.
\newblock In \emph{Proceedings of the 11th Annual Conference of the
  International Speech Communication Association}, pages 1045--1048, 2010.

\bibitem[Sundermeyer et~al.(2012)Sundermeyer, Schlüter, and
  Ney]{Sundermeyer_lstmneural}
Martin Sundermeyer, Ralf Schlüter, and Hermann Ney.
\newblock {L}{S}{T}{M} neural networks for language modeling.
\newblock In \emph{INTERSPEECH}, pages 194--197, 2012.

\bibitem[Peters et~al.(2018)Peters, Neumann, Iyyer, Gardner, Clark, Lee, and
  Zettlemoyer]{peters2018contextualized}
Matthew~E. Peters, Mark Neumann, Mohit Iyyer, Matt Gardner, Christopher Clark,
  Kenton Lee, and Luke Zettlemoyer.
\newblock Deep contextualized word representations.
\newblock \emph{arXiv preprint arXiv:1802.05365}, 2018.

\bibitem[Vaswani et~al.(2017)Vaswani, Shazeer, Parmar, Uszkoreit, Jones, Gomez,
  Kaiser, and Polosukhin]{transformer}
Ashish Vaswani, Noam Shazeer, Niki Parmar, Jakob Uszkoreit, Llion Jones,
  Aidan~N. Gomez, undefinedukasz Kaiser, and Illia Polosukhin.
\newblock Attention is all you need.
\newblock In \emph{Proceedings of the 31st International Conference on Neural
  Information Processing Systems}, NIPS'17, page 6000–6010, Red Hook, NY,
  USA, 2017.

\bibitem[Radford et~al.(2018)Radford, Narasimhan, Salimans, and
  Sutskever]{Radford2018ImprovingLU}
Alec Radford, Karthik Narasimhan, Tim Salimans, and Ilya Sutskever.
\newblock Improving language understanding by generative pre-training, 2018.

\bibitem[Howard and Ruder(2018)]{howard-ruder-2018-universal}
Jeremy Howard and Sebastian Ruder.
\newblock Universal language model fine-tuning for text classification.
\newblock In \emph{Proceedings of the 56th Annual Meeting of the Association
  for Computational Linguistics (Volume 1: Long Papers)}, pages 328--339,
  Melbourne, Australia, July 2018. Association for Computational Linguistics.

\bibitem[Devlin et~al.(2019)Devlin, Chang, Lee, and
  Toutanova]{devlin-2019-bert}
Jacob Devlin, Ming-Wei Chang, Kenton Lee, and Kristina Toutanova.
\newblock {BERT}: Pre-training of deep bidirectional transformers for language
  understanding.
\newblock In \emph{Proceedings of the 2019 Conference of the North {A}merican
  Chapter of the Association for Computational Linguistics: Human Language
  Technologies, Volume 1 (Long and Short Papers)}, pages 4171--4186,
  Minneapolis, Minnesota, June 2019. Association for Computational Linguistics.

\bibitem[Yang et~al.(2019)Yang, Dai, Yang, Carbonell, Salakhutdinov, and
  Le]{Yang2019}
Zhilin Yang, Zihang Dai, Yiming Yang, Jaime Carbonell, Russ~R Salakhutdinov,
  and Quoc~V Le.
\newblock {X}{L}{N}et: Generalized autoregressive pretraining for language
  understanding.
\newblock In H.~Wallach, H.~Larochelle, A.~Beygelzimer, F.~d\textquotesingle
  Alch\'{e}-Buc, E.~Fox, and R.~Garnett, editors, \emph{Advances in Neural
  Information Processing Systems}, volume~32, pages 5753--5763, 2019.

\bibitem[Radford et~al.(2019)Radford, Wu, Child, Luan, Amodei, and
  Sutskever]{radford2019language}
Alec Radford, Jeffrey Wu, Rewon Child, David Luan, Dario Amodei, and Ilya
  Sutskever.
\newblock Language models are unsupervised multitask learners, 2019.

\bibitem[Sun et~al.(2019)Sun, Wang, Li, Feng, Chen, Zhang, Tian, Zhu, Tian, and
  Wu]{sun2019ernie}
Yu~Sun, Shuohuan Wang, Yukun Li, Shikun Feng, Xuyi Chen, Han Zhang, Xin Tian,
  Danxiang Zhu, Hao Tian, and Hua Wu.
\newblock Ernie: Enhanced representation through knowledge integration.
\newblock \emph{arXiv preprint arXiv:1904.09223}, 2019.

\bibitem[Turing-NLG(2020)]{turingNLG}
Turing-NLG.
\newblock A 17-billion-parameter language model by {M}icrosoft, 2020.
\newblock URL
  \url{https://www.microsoft.com/en-us/research/blog/turing-nlg-a-17-billion-parameter-language-model-by-microsoft/}.

\bibitem[{White House Office of Science and Technology Policy
  (OSTP)}(2019)]{OSTP}
{White House Office of Science and Technology Policy (OSTP)}.
\newblock Guidance for regulation of artificial intelligence applications,
  2019.
\newblock URL
  \url{https://www.whitehouse.gov/wp-content/uploads/2020/01/Draft-OMB-Memo-on-Regulation-of-AI-1-7-19.pdf}.

\bibitem[Feldman(2020)]{feldman2020}
Vitaly Feldman.
\newblock Does learning require memorization? {A} short tale about a long tail.
\newblock In \emph{Proceedings of the 52nd Annual ACM SIGACT Symposium on
  Theory of Computing}, STOC 2020, page 954–959, New York, NY, USA, 2020.

\bibitem[Brown et~al.(2020{\natexlab{b}})Brown, Bun, Feldman, Smith, and
  Talwar]{brown2020memorization}
Gavin Brown, Mark Bun, Vitaly Feldman, Adam Smith, and Kunal Talwar.
\newblock When is memorization of irrelevant training data necessary for
  high-accuracy learning?
\newblock \emph{arXiv preprint arXiv:2012.06421}, 2020{\natexlab{b}}.

\bibitem[Petroni et~al.(2019)Petroni, Rockt{\"a}schel, Riedel, Lewis, Bakhtin,
  Wu, and Miller]{petroni2019}
Fabio Petroni, Tim Rockt{\"a}schel, Sebastian Riedel, Patrick Lewis, Anton
  Bakhtin, Yuxiang Wu, and Alexander Miller.
\newblock Language models as knowledge bases?
\newblock In \emph{Proceedings of the 2019 Conference on Empirical Methods in
  Natural Language Processing and the 9th International Joint Conference on
  Natural Language Processing (EMNLP-IJCNLP)}, pages 2463--2473, Hong Kong,
  China, November 2019. Association for Computational Linguistics.

\bibitem[Newman(2005)]{power_laws}
M.E.J. Newman.
\newblock Power laws, {P}areto distributions and {Z}ipf's law.
\newblock \emph{Contemporary Physics}, 46\penalty0 (5):\penalty0 323--351,
  2005.

\bibitem[Sweeney(2002)]{sweeney2002k}
Latanya Sweeney.
\newblock k-anonymity: A model for protecting privacy.
\newblock \emph{International Journal of Uncertainty, Fuzziness and
  Knowledge-Based Systems}, 10\penalty0 (05):\penalty0 557--570, 2002.

\bibitem[Al-Rfou et~al.(2016)Al-Rfou, Pickett, Snaider, Sung, Strope, and
  Kurzweil]{alrfou2016conversational}
Rami Al-Rfou, Marc Pickett, Javier Snaider, Y.-H. Sung, Brian Strope, and Ray
  Kurzweil.
\newblock Conversational contextual cues: The case of personalization and
  history for response ranking, 2016.

\bibitem[McMahan et~al.(2018)McMahan, Ramage, Talwar, and Zhang]{dprnn}
Brendan McMahan, Daniel Ramage, Kunal Talwar, and Li~Zhang.
\newblock Learning differentially private recurrent language models.
\newblock In \emph{International Conference on Learning Representations
  (ICLR)}, 2018.

\bibitem[Chelba et~al.(2013)Chelba, Mikolov, Schuster, Ge, Brants, Koehn, and
  Robinson]{billionword}
Ciprian Chelba, Tomas Mikolov, Mike Schuster, Qi~Ge, Thorsten Brants, Phillipp
  Koehn, and Tony Robinson.
\newblock One billion word benchmark for measuring progress in statistical
  language modeling, 2013.
\newblock URL \url{http://arxiv.org/abs/1312.3005}.

\bibitem[Tramèr and Boneh(2020)]{tramer2020differentially}
Florian Tramèr and Dan Boneh.
\newblock Differentially private learning needs better features (or much more
  data).
\newblock \emph{arXiv preprint arXiv:2011.11660}, 2020.

\bibitem[Papernot et~al.(2020)Papernot, Chien, Song, Thakurta, and
  Erlingsson]{papernot2020}
Nicolas Papernot, Steve Chien, Shuang Song, Abhradeep Thakurta, and Ulfar
  Erlingsson.
\newblock Making the shoe fit: Architectures, initializations, and tuning for
  learning with privacy, 2020.
\newblock URL \url{https://openreview.net/forum?id=rJg851rYwH}.

\bibitem[Gokaslan and Cohen(2019)]{Gokaslan2019OpenWeb}
Aaron Gokaslan and Vanya Cohen.
\newblock Openwebtext corpus.
\newblock \url{http://Skylion007.github.io/OpenWebTextCorpus}, 2019.

\bibitem[Merity et~al.(2017)Merity, Xiong, Bradbury, and Socher]{wikitext}
Stephen Merity, Caiming Xiong, James Bradbury, and Richard Socher.
\newblock Pointer sentinel mixture models.
\newblock \emph{ICLR}, 2017.

\bibitem[Zhang et~al.(2017)Zhang, Bengio, Hardt, Recht, and Vinyals]{zhang2016}
Chiyuan Zhang, Samy Bengio, Moritz Hardt, Benjamin Recht, and Oriol Vinyals.
\newblock Understanding deep learning requires rethinking generalization.
\newblock \emph{ICLR}, 2017.

\bibitem[Zanella-B\'{e}guelin et~al.(2020)Zanella-B\'{e}guelin, Wutschitz,
  Tople, R\"{u}hle, Paverd, Ohrimenko, K\"{o}pf, and
  Brockschmidt]{snapshotattack}
Santiago Zanella-B\'{e}guelin, Lukas Wutschitz, Shruti Tople, Victor R\"{u}hle,
  Andrew Paverd, Olga Ohrimenko, Boris K\"{o}pf, and Marc Brockschmidt.
\newblock Analyzing information leakage of updates to natural language models.
\newblock In \emph{Proceedings of the 2020 ACM SIGSAC Conference on Computer
  and Communications Security}, page 363–375, 2020.

\bibitem[Shokri et~al.(2017)Shokri, Stronati, Song, and Shmatikov]{shokri17}
Reza Shokri, Marco Stronati, Congzheng Song, and Vitaly Shmatikov.
\newblock Membership inference attacks against machine learning models.
\newblock In \emph{2017 IEEE Symposium on Security and Privacy (SP)}, pages
  3--18, 2017.

\bibitem[Yeom et~al.(2018)Yeom, Giacomelli, Fredrikson, and Jha]{yeom18}
Samuel Yeom, Irene Giacomelli, Matt Fredrikson, and Somesh Jha.
\newblock Privacy risk in machine learning: Analyzing the connection to
  overfitting.
\newblock In \emph{2018 IEEE 31st Computer Security Foundations Symposium
  (CSF)}, pages 268--282, 2018.

\bibitem[Song and Shmatikov(2019)]{song19}
Congzheng Song and Vitaly Shmatikov.
\newblock Auditing data provenance in text-generation models.
\newblock In \emph{KDD}, 2019.

\bibitem[Nasr et~al.(2019)Nasr, Shokri, and Houmansadr]{nasr19}
Milad Nasr, Reza Shokri, and Amir Houmansadr.
\newblock Comprehensive privacy analysis of deep learning: Passive and active
  white-box inference attacks against centralized and federated learning.
\newblock In \emph{2019 IEEE Symposium on Security and Privacy (SP)}, pages
  739--753, 2019.

\bibitem[Long et~al.(2018)Long, Bindschaedler, Wang, Bu, Wang, Tang, Gunter,
  and Chen]{long18}
Yunhui Long, Vincent Bindschaedler, Lei Wang, Diyue Bu, Xiaofeng Wang, Haixu
  Tang, Carl~A. Gunter, and Kai Chen.
\newblock Understanding membership inferences on well-generalized learning
  models.
\newblock \emph{arXiv preprint arXiv:1802.04889}, 2018.

\bibitem[Hayes et~al.(2019)Hayes, Melis, Danezis, and Cristofaro]{LOGAN}
Jamie Hayes, Luca Melis, George Danezis, and Emiliano~De Cristofaro.
\newblock {LOGAN}: Membership inference attacks against generative models.
\newblock \emph{Proceedings on Privacy Enhancing Technologies}, 2019\penalty0
  (1):\penalty0 133 -- 152, 2019.

\bibitem[Truex et~al.(2018)Truex, Liu, Gursoy, Yu, and Wei]{truex18}
Stacey Truex, Ling Liu, Mehmet~Emre Gursoy, Lei Yu, and Wenqi Wei.
\newblock Towards demystifying membership inference attacks.
\newblock \emph{arXiv preprint arXiv:1807.09173}, 2018.

\bibitem[Irolla and Châtel(2019)]{irolla19}
Paul Irolla and Grégory Châtel.
\newblock Demystifying the membership inference attack.
\newblock In \emph{2019 12th CMI Conf. on Cybersecurity and Privacy (CMI)},
  pages 1--7, 2019.

\bibitem[Hisamoto et~al.(2020)Hisamoto, Post, and Duh]{hisamoto20}
Sorami Hisamoto, Matt Post, and Kevin Duh.
\newblock Membership inference attacks on sequence-to-sequence models: Is my
  data in your machine translation system?
\newblock \emph{TACL}, 8:\penalty0 49--63, 2020.

\bibitem[Salem et~al.(2018)Salem, Zhang, Humbert, Fritz, and Backes]{salem18}
Ahmed Salem, Yang Zhang, Mathias Humbert, Mario Fritz, and Michael Backes.
\newblock {M}{L}-{L}eaks: Model and data independent membership inference
  attacks and defenses on machine learning models.
\newblock \emph{arXiv preprint arXiv:1806.01246}, 2018.

\bibitem[Sablayrolles et~al.(2019)Sablayrolles, Douze, Schmid, Ollivier, and
  Jegou]{sablayrolles19a}
Alexandre Sablayrolles, Matthijs Douze, Cordelia Schmid, Yann Ollivier, and
  Herve Jegou.
\newblock White-box vs black-box: {B}ayes optimal strategies for membership
  inference.
\newblock In \emph{Proceedings of the 36th International Conference on Machine
  Learning}, volume~97 of \emph{Proceedings of Machine Learning Research},
  pages 5558--5567, Long Beach, California, USA, 09--15 Jun 2019.

\bibitem[Leino and Fredrikson()]{Leino20}
Klas Leino and Matt Fredrikson.
\newblock Stolen {M}emories: Leveraging model memorization for calibrated
  white-box membership inference.
\newblock \emph{29th USENIX Security Symposium 2020}.

\bibitem[Choquette-Choo et~al.(2020)Choquette-Choo, Tramer, Carlini, and
  Papernot]{choo2020labelonly}
Christopher~A. Choquette-Choo, Florian Tramer, Nicholas Carlini, and Nicolas
  Papernot.
\newblock Label-only membership inference attacks.
\newblock \emph{arXiv preprint arXiv:2007.14321}, 2020.

\bibitem[Murakonda and Shokri(2020)]{murakonda2020ml}
Sasi~Kumar Murakonda and Reza Shokri.
\newblock {M}{L} {P}rivacy {M}eter: Aiding regulatory compliance by quantifying
  the privacy risks of machine learning.
\newblock \emph{arXiv preprint arXiv:2007.09339}, 2020.

\bibitem[Kairouz et~al.(2019)Kairouz, McMahan, Avent, Bellet, Bennis, Bhagoji,
  Bonawitz, Charles, Cormode, Cummings, and et~al.]{kairouz19}
Peter Kairouz, H.~Brendan McMahan, Brendan Avent, Aur{\'{e}}lien Bellet, Mehdi
  Bennis, Arjun~Nitin Bhagoji, Keith Bonawitz, Zachary Charles, Graham Cormode,
  Rachel Cummings, and et~al.
\newblock Advances and open problems in federated learning.
\newblock \emph{arXiv preprint arXiv:1912.04977}, 2019.

\end{thebibliography}
\bibliographystyle{unsrtnat}

\appendix
\section{Complexity of our approach}
\label{sec:complexity}
We utilize Spark for an efficient implementation of our approach. The model evaluation was the computation bottleneck stage in the pipeline, since Spark does not offer tools to run deep neural networks efficiently. To mitigate this and maximise the Spark clusters' CPU utilization during model evaluation, we batch all the rows in each Spark RDD partition as one batched tensor and feed it to the model. Subsequently, we re-partition and convert the output to a dataframe for the next stage. This method allows us to achieve a sub-linear improvement of the average processing speed per line as a function of the number of Spark nodes. Consequently, by increasing the number nodes linearly with the number of rows in the database we achieve approximately $O(1)$ complexity.


\section{Tab attack for the DP-LM RanIni $\epsilon = \cdot$ models in Section \ref{sec:reddit}}
\label{sec:a2}
We present the leakage analysis for the unique sequences coming out of the tab attack for the DP-LM RanIni $\epsilon = \cdot$ models in Table \ref{table:eps_4_68}, \ref{table:eps_6_20}, and \ref{table:eps_26_4}.
\begin{table*}[ht]
\caption{Unique sequences from the tab attack for the DP-LM RanIni $\epsilon = 4.68$ model in Section \ref{sec:reddit}.}
\label{table:eps_4_68}
\vskip 0.15in
\begin{center}
\begin{small}
\begin{sc}
\begin{tabular}{p{0.4\textwidth}p{0.11\textwidth}p{0.1\textwidth}p{0.14\textwidth}}
\toprule
$\mathcal{S}_{\textnormal{uniq}}$ & total \# in $\mathcal{D}$ & user \# in $\mathcal{D}$ & context len \\
\midrule
``way , I don't think it is" & 1 & 1 & 2  \\
``the time , I would be" & 1 & 1 & 8 \\
``same thing , I would be" & 1 & 1 & 10 \\
``media ) is not" & 1 & 1 & 10 \\
``not be ) but" & 1 & 1 & 9 \\
\bottomrule
\end{tabular}
\end{sc}
\end{small}
\end{center}
\vskip -0.1in
\end{table*}

\begin{table*}[ht]
\caption{Unique sequences from the tab attack for the DP-LM RanIni $\epsilon = 6.20$ model in Section \ref{sec:reddit}. The $\textnormal{PP}_{\textnormal{lm}}(\cdot)$ column is the perplexity of each sequence with respect to the DP-LM RanIni $\epsilon = 6.20$ model. The $\textnormal{PP}_{\textnormal{public}}(\cdot)$ column is the perplexity of each sequence with respect to the public model. The last column is the log-ratio of the perplexities of the previous two columns. The worst-case leakage epsilon is $\epsilon_l = 0.64$. We refer the following model as the public model in this table. We remove the users that are the owners of these unique sequences and all of their data (not just these sequences) from the training data and train a new model with the remaining users. We consider the new model as a public model for the users of these unique sequences since it has never seen any data of these users during its training.}
\label{table:eps_6_20}
\vskip 0.15in
\begin{center}
\begin{small}
\begin{sc}
\begin{tabular}{p{0.25\textwidth}p{0.11\textwidth}p{0.1\textwidth}p{0.12\textwidth}p{0.05\textwidth}p{0.065\textwidth}p{0.1\textwidth}}
\toprule
$\mathcal{S}_{\textnormal{uniq}}$ & total \# in $\mathcal{D}$ & user \# in $\mathcal{D}$ & context len & $\textnormal{PP}_{\textnormal{lm}}(\cdot)$ & $\textnormal{PP}_{\textnormal{public}}(\cdot)$ & $\log \dfrac{\textnormal{PP}_{\textnormal{public}}(\cdot)}{\textnormal{PP}_{\textnormal{lm}}(\cdot)}$\\ 
\midrule 
``said , I think you should be able to" & 1 & 1 & 3 & 4.39 & 5.21 & 0.17  \\
``me a link to your post ?" & 1 & 1 & 4 & 3.33 & 3.49 & 0.05\\
``you feel better , then you can" & 1 & 1 & 4 & 5.4 & 5.53 & 0.02 \\
``has any questions or concerns , please" & 1 & 1 & 5 & 5.1 & 8.87 & 0.55 \\
``as I know , I think the" & 1 & 1 & 3 & 3.75 & 4.5 & 0.18 \\
``Court , he would have" & 1 & 1 & 8 & 4.46 & 5.22 & 0.16 \\
``want * to be ?" & 1 & 1 & 13 & 4.17 & 3.63 & -0.14 \\
``like is that you are" & 1 & 1 & 4 & 5.35 & 5.78 & 0.08 \\
``of people ) are" & 1 & 1 & 6 & 5.14 & 4.21 & -0.2 \\
``Wars , we have" & 1 & 1 & 5 & 7.28 & 7.36 & 0.01 \\
``Wars ) is" & 1 & 1 & 7 & 3.53 & 6.69 & \textbf{0.64} \\
\bottomrule
\end{tabular}
\end{sc}
\end{small}
\end{center}
\vskip -0.1in
\end{table*}

\begin{table*}[ht]
\caption{Unique sequences from the tab attack for the DP-LM RanIni $\epsilon = 26.4$ model in Section \ref{sec:reddit}.}
\label{table:eps_26_4}
\vskip 0.15in
\begin{center}
\begin{small}
\begin{sc}
\begin{tabular}{p{0.4\textwidth}p{0.11\textwidth}p{0.1\textwidth}p{0.14\textwidth}p{0.08\textwidth}}
\toprule
$\mathcal{S}_{\textnormal{uniq}}$ & total \# in $\mathcal{D}$ & user \# in $\mathcal{D}$ & context len \\
\midrule
``the other hand , I would have to" & 1 & 1 & 2   \\
``lot of people who don't know ." & 1 & 1 & 12  \\
``the case , then I would have" & 1 & 1 & 4   \\
``fair , I don't think you can" & 1 & 1 & 3   \\
``idea what you are doing , you" & 1 & 1 & 5  \\
``as I know , I have a" & 1 & 1 & 3   \\
``to do , I would have to" & 1 & 1 & 10   \\
``you feel better , then you can" & 1 & 1 & 4   \\
``your question , you can only" & 1 & 1 & 3   \\
``of curiosity , I think the" & 1 & 1 & 2   \\
``idea what he was doing ?" & 1 & 1 & 6   \\
``doing , you can get a" & 1 & 1 & 7   \\
``Court have to do with the" & 1 & 1 & 7   \\
``point , it is not ." & 1 & 1 & 14   \\
``different situation , it's a" & 1 & 1 & 6   \\
``of years , I was" & 1 & 1 & 4   \\
``change , they would be" & 1 & 1 & 6   \\
``assault , I would be" & 1 & 1 & 12   \\
``* are * really *" & 1 & 1 & 8   \\
``* do it , then" & 1 & 1 & 4  \\
``media ) is not" & 1 & 1 & 10   \\
``Court , then they" & 1 & 1 & 7   \\
``butter , I would" & 1 & 1 & 7   \\
``Francisco , I think" & 1 & 1 & 6   \\
``Court have to be" & 1 & 1 & 5   \\
``of people ) are" & 1 & 1 & 6   \\
``\% agree )." & 1 & 1 & 15  \\
``Arabia * is" & 1 & 1 & 8   \\
``Arabia ) is" & 1 & 1 & 7   \\
``Wars ) is" & 1 & 1 & 7   \\
\bottomrule
\end{tabular}
\end{sc}
\end{small}
\end{center}
\vskip -0.1in
\end{table*}

\end{document}